\documentclass[runningheads]{llncs}
\usepackage[T1]{fontenc}
\usepackage[sort]{cite}
\usepackage{hyperref}
\usepackage{graphicx}
\begin{document}
\title{Exploring LLM-Generated Feedback for Economics Essays: How Teaching Assistants Evaluate and Envision Its Use}
%
\titlerunning{Exploring LLM-Generated Feedback for Economics Essays}
%
\author{Xinyi Lu, Aditya Mahesh$^*$, Zejia Shen$^*$, Mitchell Dudley, Larissa Sano,\\ Xu Wang}
\def\thefootnote{*}\footnotetext{Both authors contributed equally to this work}
\authorrunning{X. Lu et al.}
%
\institute{University of Michigan, Ann Arbor MI 48109, USA\\ 
\email{\{lwlxy, mahesha, jasnshen, mrdudley, llubomud, xwanghci\}@umich.edu}}
%
\maketitle              
\vspace{-1.5pc}
\begin{abstract}
This project examines the prospect of using AI-generated feedback as suggestions to expedite and enhance human instructors’ feedback provision. In particular, we focus on understanding the teaching assistants’ perspectives on the quality of AI-generated feedback and how they may or may not utilize AI feedback in their own workflows.  
We situate our work in a foundational college Economics class, which has frequent short essay assignments. We developed an LLM-powered feedback engine that generates feedback on students’ essays based on grading rubrics used by the teaching assistants (TAs). 
To ensure that TAs can meaningfully critique and engage with the AI feedback, we had them complete their regular grading jobs. For a randomly selected set of essays that they had graded, we used our feedback engine to generate feedback and displayed the feedback as in-text comments in a Word document. We then performed think-aloud studies with 5 TAs over 20 1-hour sessions to have them evaluate the AI feedback, contrast the AI feedback with their handwritten feedback, and share how they envision using the AI feedback if they were offered as suggestions. 
The study highlights the importance of providing detailed rubrics for AI to generate high-quality feedback for knowledge-intensive essays. TAs considered that using AI feedback as suggestions during their grading could expedite grading, enhance consistency, and improve overall feedback quality. We discuss the importance of decomposing the feedback generation task into steps and presenting intermediate results, in order for TAs to use the AI feedback. 

\vspace{-1pc}
\end{abstract}
\vspace{-2.75pc}
\section{Introduction}
\vspace{-0.75pc}
Extensive research has shown that feedback is important for learning \cite{hattie2007power,hattie2018visible,chi2014icap,hicks2016framing,koedinger2012knowledge}, yet high quailty feedback requires expertise and efforts to write \cite{nelson2009nature,patchan2016nature}.
Since the rise of generative AI, research communities around AI and Education have explored using large language models (LLMs) to generate tutoring responses and feedback. 
A number of studies showed promising results that when instructed well, LLMs can generate high quality feedback that is similar to human feedback \cite{jia2024llm,heickal2024generating,wang2024bridging,dai2023can}. However, many studies observe problems in LLM feedback such as they can be too general \cite{jia2024llm}, cannot capture nuanced differences in students' answers \cite{heickal2024generating,jia2024llm,steiss2024comparing}, hallucinate and make mistakes \cite{jia2024llm}. In this work, we aim to address the question: Even when AI feedback is imperfect, can it be used as suggestions to expedite and enhance human instructors' feedback provision? 


This work aims to address key gaps in our understanding of LLMs' capabilities in generating effective feedback. 
First, prior work on LLM-based feedback generation has primarily focused on single-turn, single-rubric evaluations of short answers in subjects such as math and programming. Several studies have examined using LLMs to generate tutoring moves for math problems \cite{wang2024bridging,scarlatos2024improving,mcnichols2024can,heickal2024generating}, while others have examined their use in providing feedback to human tutors on tutoring strategies, such as encouraging praise \cite{kakarla2025comparing,xu2025improving}. Some research has investigated feedback generation for longer texts, including essays for English learners \cite{escalante2023ai,steiss2024comparing} and project reports \cite{jia2024llm}. However, these studies primarily assess language features, with limited exploration of LLMs' effectiveness in evaluating knowledge accuracy in longer essays. To address this gap, this work examines LLM-generated feedback for knowledge-intensive essays in a college-level introductory economics class, where students write essays to explain economic concepts and phenomena. 
Second, existing approaches have focused on developing techniques and pipelines to align LLM-generated feedback with human feedback using quantitative metrics such as accuracy, recall, and linguistic overlap \cite{almegren2024evaluating,latif2024fine}. In contrast, this study investigates the potential for human-AI collaboration, exploring whether AI-generated feedback -- despite its imperfections -- can enhance instructors' grading experiences. 
Third, generating feedback for a whole essay presents unique challenges beyond short-answer evaluation, particularly in the context of human-AI interaction. Prior research has shown that instructors prefer to critically review AI-generated content before using it, and seek to understand the rationale behind the AI-generated content \cite{lu2023readingquizmaker}. This work also examines the UI challenges of visualizing AI-generated feedback within essays, aiming to enhance usability and instructor trust. 

We conducted our study in a college-level Introduction to Economics (ECON101) course. It has frequent short-essay assignments, such as "identify an economic phenomenon involving market failure and analyze the market failure in it". We consider these as knowledge-intensive essays, as they require students to demonstrate a precise understanding of economic concepts through their writing. These assignments also have well-defined rubrics, e.g., ``correctly explain the definition of market failure: the market fails to allocate an efficient quantity without government intervention.'' Extensive prior research has shown that rubrics improve feedback quality \cite{patchan2016nature,yuan2016almost}, a finding that also holds when using LLMs to generate feedback \cite{steiss2024comparing,xu2025improving,scarlatos2024improving,jia2024llm,wang2024bridging}. Given this, we chose this course --ECON101-- with well-established rubrics to examine an ideal scenario: When detailed rubrics are provided, how does LLM provide feedback on a knowledge-intensive essay?

We propose a feedback engine that decomposes the feedback-generation task into three subtasks. For each rubric, the feedback engine follows these steps: 1) AI identifies sentences in the essay relevant to the rubric; 2) AI makes a judgment on whether the essay satisfies this rubric; 3) AI generates a feedback message to guide the student toward achieving the rubric without explicitly providing the correct answer.
We conducted the study in Fall 2024 in ECON101, where teaching assistants (TAs) provided feedback on students' essays. 
To create an authentic environment for TAs to critically compare their feedback with the AI's feedback, we first asked TAs to complete their grading tasks as usual. After grading, we invited them to participate in think-aloud sessions where they reviewed AI-generated feedback on the same set of graded essays. During these sessions, TAs compared their feedback with the AI’s and reflected on whether they would incorporate AI-generated suggestions into their workflow. The study spanned 4 writing assignments, with 5 TAs participating in a total of 20 one-hour sessions.

Through this study, we aimed to answer the following research questions: 1) How do TAs perceive the differences between AI-generated feedback and their own feedback? 
2) What are the prospects of using AI-generated feedback as suggestions in the TA grading process? Specifically, can AI feedback be helpful and how should it be presented to enhance grading effectiveness?

Here is a summary of our findings: 1) AI feedback exhibits more characteristics of effective feedback than human feedback, including the use of praise, explanations, and guiding questions. 2) AI-generated feedback aligns more closely with the rubrics, which can be a double-edged sword-- it ensures consistency but may be misleading when the AI applies the rubric too rigidly. 3) AI feedback can be overly fragmented due to rubric-based structuring. This may overlook the need for holistic evaluations that could better support students' learning. However, for knowledge-intensive essays, AI cannot generate accurate feedback without detailed rubrics. 4) Despite its errors, TAs found AI mistakes easy to correct, especially when intermediate AI outputs—such as in-text highlights—were visualized. TAs responded positively to a human-AI collaborative approach, where AI-generated feedback serves as suggestions. They viewed this approach as a way to expedite grading, enhance consistency, and improve overall feedback quality.

\vspace{-1pc}
\section{Related Work}
\vspace{-0.5pc}
\subsection{Existing work on evaluating AI feedback}
\vspace{-0.5pc}
Recent work has shown the potential of using AI to generate feedback comparable to that of human experts' \cite{almegren2024evaluating,latif2024fine,escalante2023ai}. Researchers showed that AI-generated feedback could have higher coherence than human-crafted ones \cite{dai2023can}, and maintains objectivity \cite{Almasre2024DevelopmentAE}
across large volumes of evaluations.
However, these objective metrics lack insights from both the experts and the students. Jia et al. explored students' and teachers' perspectives on AI feedback. They found limitations in AI feedback including hallucination and vague content, and suggested the irreplaceable nature of human feedback \cite{jia2024llm}. 
In this project, we aim to explore the potential for human-AI collaboration in providing feedback, and investigate whether AI-generated feedback can assist with the process despite its limitations.

\vspace{-1pc}
\subsection{Capabilities of LLMs in modular tasks}
\vspace{-0.5pc}
Recent work has shown the capacity of LLMs in a wide range of modular natural language tasks. They demonstrated strong ability in retrieving relevant content and producing highly fluent text. Moreover, with advanced prompting techniques like Chain-of-Thought (CoT) \cite{wei2022chain}, LLMs show notable reasoning abilities and produce rational decisions \cite{eigner2402determinants}. However, LLMs are limited by their lack of domain-specific knowledge. And even worse, when they lack relevant knowledge, instead of expressing uncertainty or declining to respond, they often generate inaccurate content, leading to hallucinations, especially for complex tasks \cite{lewis2020retrieval}. One approach to mitigate hallucinations is to improve AI explainability \cite{luo2024understanding,ji2023survey}. Researchers developed algorithms leveraging explainability to improve truthfulness in generations \cite{li2024inference}. Other work showed that exposing the source of the generated text helps the users better identify hallucinations \cite{rashkin2023measuring}. In this work, instead of using LLM to perform a single-turn, end-to-end feedback generation, we decompose the task into such modular sub-tasks that LLM has shown high performance on and are easy to validate. To better support users in evaluating the generated content, we present the results from all the intermediate steps.
\vspace{-1pc}
\section{Method}
\vspace{-0.5pc}
\subsection{Tasks}
\vspace{-0.5pc}
We conducted our study in a college-level Introduction to Economics (ECON101) course, which has frequent knowledge-intensive essay assignments requiring a precise understanding of economic concepts. These assignments also have well-defined rubrics. Additionally, the course instructors provided historical feedback in response to each rubric, which is a feedback message. And it is suggested to provide to the students, if the student doesn't meet this rubric. 

\vspace{-1pc}
\subsection{Feedback generation and visualization}
\begin{figure}
    \centering
    \includegraphics[width=0.95\linewidth]{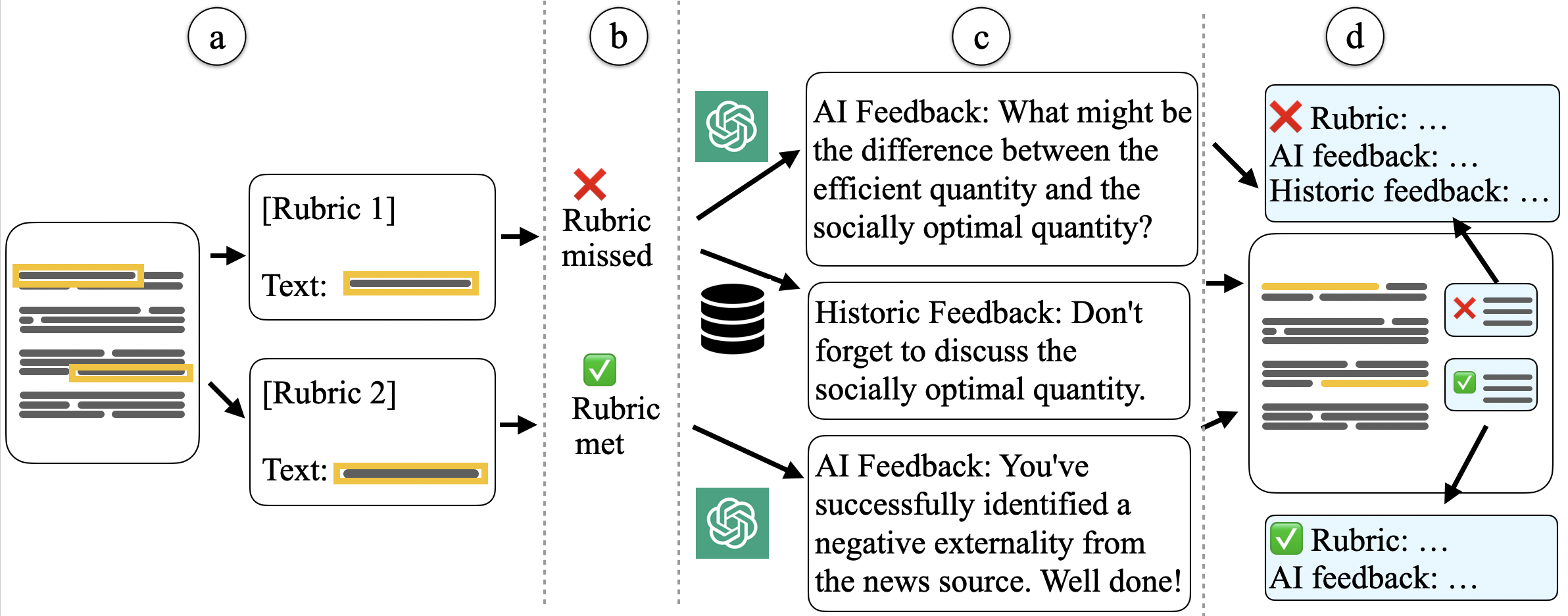}
    \vspace{-1pc}
    \caption{The feedback generation is decomposed into steps. With the relevant sentences for each rubric identified in (a), AI makes judgments on whether each rubric is met (b). Then the relevant sentences and judgments are used to generate feedback (c). Additionally, if the rubric is missed, historic feedback is retrieved from a set of feedback designed by instructors. The feedback and relevant information will be shown as in-text comments in a Word document (d).}
    
    \label{fig:diagram}
    \vspace{-1pc}
\end{figure}

\vspace{-0.5pc}
Instead of adopting an end-to-end generation for feedback, we decomposed the task and developed an LLM-powered pipeline, as shown in Figure \ref{fig:diagram}. Specifically, the pipeline generates a piece of feedback for each rubric separately. To improve the localization of each feedback, we adopted the idea of Chain-of-Thought \cite{wei2022chain} and structured the process into three steps. First, we requested the AI model to identify the most relevant sentences in the student's essay that indicate whether the response meets or misses the rubric. These sentences serve as evidence for the subsequent judgment. Next, based on this judgment, the model generates feedback while explicitly providing a rationale.

To better understand the preferable form of AI assistance, we explored two sets of AI-assisted feedback paradigms. In one variant, AI is only used to make judgments, which are used to retrieve the historic feedback. 

In the other variant, the feedback is generated entirely by AI. To produce feedback aligned with high-quality feedback guidelines, we specified the suggestions by Patchan et al. \cite{patchan2016nature} in the system prompt for feedback generation. Specifically, it is specified to (1) use specific and localized language in all feedback; (2) provide praise when the student meets the rubrics; and (3) if the student doesn't meet the rubrics, pinpoint the student's mistake and pose guiding questions without revealing the answer. Additionally, following the suggestion of few-shot learning \cite{brown2020language}, examples of high-quality feedback created by the instructor were also provided. 
All the feedback was generated by GPT-4o with a temperature of 0.05 to enhance consistency. \footnote{\href{https://github.com/UM-Lifelong-Learning-Lab/AIED2025-Exploring-LLM-Generated-Feedback-for-Economics-Essay}{https://github.com/UM-Lifelong-Learning-Lab/AIED2025-Exploring-LLM-Generated-Feedback-for-Economics-Essay}}

\begin{figure}
    \vspace{-1.5pc}
    \centering
    \includegraphics[width=\linewidth]{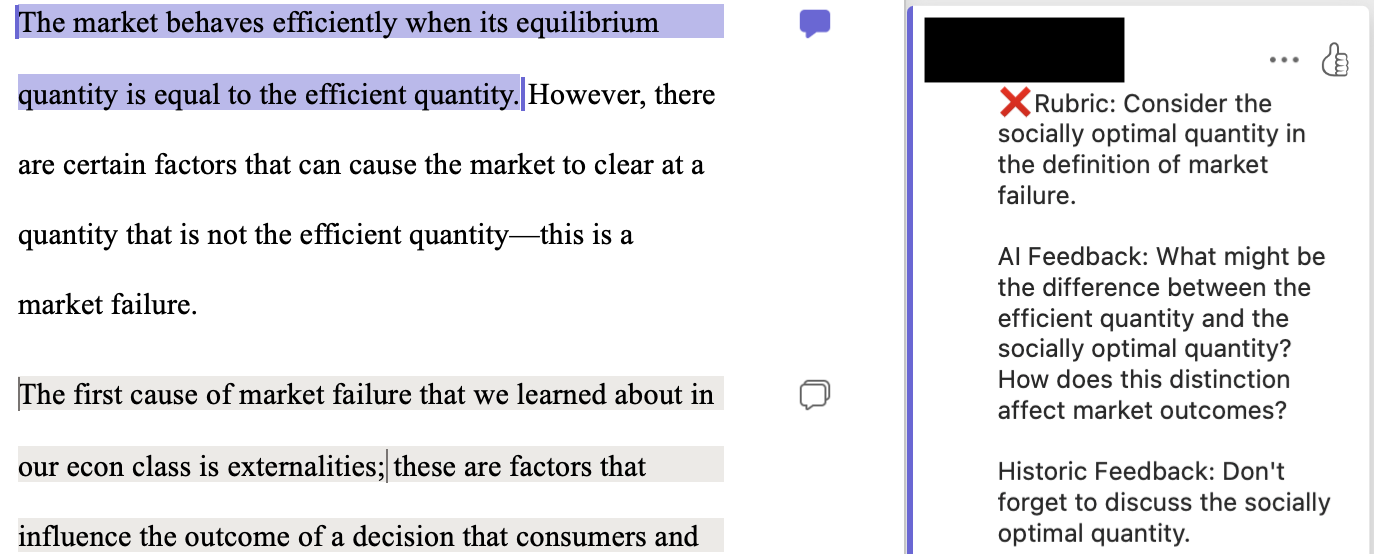}
    \vspace{-1.5pc}
    \caption{AI-generated feedback is shown as in-text comments on a Word document, added to the most relevant sentences in the student essay. Each comment contains the rubric, the AI judgment, the AI feedback, and the historic feedback.}
    \label{fig:comment_example}
    \vspace{-1pc}
\end{figure}

To visualize the feedback, we developed a Word plugin to integrate the generated feedback as in-text comments on a document. To improve the AI explainability of the feedback, the plugin highlighted the relevant sentences identified by AI in the generation. As shown in Figure \ref{fig:comment_example}, each comment includes the following information -- the corresponding rubric item, AI judgment, the historic feedback, and the AI feedback. This is to help users better identify hallucinations.

\vspace{-1pc}
\subsection{Study Design}
\vspace{-0.5pc}
We conducted the study in Fall 2024 in ECON101, where TAs provided feedback to students' essays. The study is IRB-approved. 
To create an authentic environment for TAs to critically compare their feedback with the AI's feedback, we first asked them to complete their grading tasks as usual. After grading, we invited the TAs to participate in think-aloud sessions where they reviewed AI-generated feedback on the same set of graded essays. During these sessions, TAs compared their feedback with the AI’s and reflected on whether they would incorporate AI-generated suggestions into their workflow. The study spanned four writing assignments, with five TAs participating in a total of 20 one-hour sessions.
Participants were compensated with a \$25 Gift Card for each study session. 

\vspace{-1pc}
\subsubsection{Procedure}
After getting participants' consent, we gave an introduction on how the AI feedback and historic feedback were generated. We explained that the feedback is generated based on the highlight and the AI judgment. Then participants were presented with a randomly selected set of essays they had graded, with AI feedback as in-text comments in a Word document as shown in Figure~\ref{fig:comment_example}. Participants were asked to evaluate the AI feedback, and contrast the AI feedback with their own feedback on the same essay. We also specifically asked about how they envision using the AI feedback if provided as they graded.
\vspace{-2pc}
\subsubsection{Data analysis methods}
The recordings were transcribed, de-identified, and analyzed with affinity diagram \cite{moggridge2007designing}. Two authors interpreted the transcripts, iteratively grouped their notes, and identified key emerging themes.
\vspace{-1pc}
\section{Findings}
\vspace{-0.5pc}
\subsection{RQ1: How do TAs perceive the differences between AI-generated feedback and their own feedback? }
\subsubsection{AI feedback exhibits more characteristics of effective feedback, such as providing praise, using guiding questions and explanations.} \label{guidelines}
First, AI provides more positive feedback (praises), which is often neglected by the TAs. Many participants mentioned that they would directly incorporate the positive feedback during grading. P5 said, \textit{``I would definitely add that (positive) comment just as it is.''} In contrast, TAs often do not prioritize leaving positive comments given time constraints in grading. As P2 explained, \textit{``Honestly, I was grading so many of these. If I had time I would have liked to put in nice comments, so I would have definitely used AI's positive comments.''}

Second, participants value AI feedback for using guiding questions as scaffolds. Many TAs agreed on the importance of providing scaffolds in their feedback when students had an opportunity to revise their work. For instance, P5 found the guiding questions helpful for \textit{``push[ing] the students in a certain direction where they're already kind of like on that page.''} At the same time, they found it challenging to craft guiding questions on their own. As P2 said, \textit{``the hardest part is not giving them the answer, but also leading them in the right direction.''} 

Third, AI feedback often contained explanations of the students' mistakes. For example, when the student mistook fish, which is a rival good, to be non-rival, the TA's feedback was, ``Fish is rival instead of non-rival''. The historic feedback was also generic: ``Discuss the concepts of rivalry in the context of the news article'', while the AI feedback provides a detailed rationale that ``... tuna, as a finite resource, is rival because one person's consumption reduces availability for others.'' P5 preferred the AI feedback, stating \textit{``I feel like overall, this comment is probably better than the comment I left ... It gives a good explanation as to why (the student was wrong).''} 
\vspace{-1pc}
\subsubsection{AI feedback is more personalized to students' responses in comparison to the historic feedback.}
For example, when the student mentioned that the nearby communities are the ``third party'', and needed to further analyze the involuntary nature of the nearby communities, AI feedback says, ``Consider explaining how the communities near the factories are involuntarily affected by pollution.'' 
On the contrary, the historic feedback provided by the instructors is designed to be more generic and may not apply to all students. As P2 said, \textit{``We had (historical) comments that we could use... Even those I'd like to tweak a little bit for the specific essays.''} Below is an example contrasting the difference between the AI and the historic feedback.

\begin{quote}
    \textbf{Rubric:} ``Explain that the thrid party in the negative externality is involuntarily affected.''\\
    \textbf{Student response: } ``In economics, we call this a negative externality; the social costs are not taken on by the producers or consumers but by society.''\\
    \textbf{AI Feedback:} ``How might the impact on individuals differ if they were voluntary participants in the market? Consider how the concept of choice plays into the definition of negative externalities.''\\
    \textbf{Historic Feedback:} ``Revisit the definition of an externality and consider how those affected are economically reflected in the market.''
\end{quote}

When presented with both the AI feedback and the historic feedback above, P4 preferred the AI feedback because it is more personalized to the student's problem saying \textit{``I like that (the AI feedback) uses the word `voluntary' here, because that's kind of the direction you're trying to point them in... (The historic feedback) is just talking about the 3rd party which the student already discusses.''}

\vspace{-1pc}
\subsubsection{AI-generated feedback aligns more closely with the rubrics, which can be a double-edged sword—it ensures consistency but may be misleading when the AI applies the rubric too rigidly.
}
Participants appreciated that the AI feedback was better aligned with the rubrics and more fine-grained since the feedback engine generates one feedback message per rubric item, while the TAs might give a combined feedback for several rubric items. However, this brings about the trade-off that AI feedback could be misleading when the rubrics are not well written. We will describe scenarios where AI tends to make mistakes.  

First, participants found that AI frequently made mistakes on assessing students' definitions of specialized economic terms, when the definitions were not provided in the rubric. For example, on the rubric item ``Correctly use the terms quantity demanded vs. demand'', AI often misjudged responses because it doesn't have the expert knowledge to differentiate ``quantity demanded'' and ``demand''.

Second, AI can be very strict about a rubric item and may reject alternative ways of writing that also satisfy the rubric criterion. 
For example, a decrease in demand can also be expressed as ``a demand curve shifts leftwards'' or ``a demand curve shifts downwards'' or ``a reduction in the consumers' willingness to pay for the good''. 
Both P4 and P5 noted that some students implicitly conveyed the intended idea in their writing without explicitly using the language in the rubric, yet AI marked them incorrect. 

Third, TAs and AI had different requirements for the depth of explanation in the students' writing.
For example, for the rubric item ``Explain the change from nonbinding to binding price floor'', the student correctly stated that the policy posed a binding minimum wage, which was marked as correct by AI. 
However, P1 expected a deeper explanation, \textit{``The binding minimum wage. That's good. But I also wanted them to describe the difference between binding and non-binding.''}  

Lastly, participants found that AI could be too stringent by focusing on unnecessary details in the rubric. 
When a rubric consists of multiple components, TAs can better identify and prioritize key ideas, whereas AI may lose focus during evaluation and flag minor points that are peripheral to the core concepts.

\vspace{-1pc}
\subsubsection{TAs’ feedback is more holistic, extending beyond the rubrics to consider broader aspects of student understanding and writing quality.}
While AI-generated feedback typically targets individual rubric items, TAs adopt a more holistic approach. They do not treat each rubric item in isolation; rather, they sometimes synthesize multiple rubrics and focus on overarching aspects such as conceptual understanding and the flow of ideas. For example, P2 said they would combine AI-generated feedback for two different rubric items to create a comprehensive feedback message.
Here, we summarize the key considerations TAs took into account when handwriting their feedback.

First, TAs shared that there were cases in which students might be struggling with deeper conceptual issues that go beyond merely missing a single rubric. In such cases, they found the AI feedback focusing on a single rubric to be insufficient. As P4 mentions, \textit{``I feel like a more specific comment (than the AI feedback or historic feedback) was needed, just because the student was so wrong.''} For instance, one rubric item instructs students to ``Mention that automation and labor are substitutes for consumption'', prompting the student to use economic terms to analyze how changes in automation affect the labor market. The corresponding AI feedback suggested ``To strengthen your argument, mention that automation and labor are substitutes in consumption. This will help explain why firms might switch to automation when labor costs increase.'' However, the student’s essay showed no recognition of the relationship between automation and labor, indicating a more fundamental misunderstanding rather than omission of terminology. Because the AI’s suggestion did not address this deeper conceptual gap, the TA found it inadequate. As P4 explained: \textit{"So I don't think this comment really helps the student understand what's going on here. I would ... try to guide them in a way to talk about how the demand for automation is shifting... I like that the AI mentioned that they should explain that they're substitutes. That's important. But this (understanding the relationship) part is also important."}

Second, some rubrics do not apply to all the essays, and some students' mistakes are not covered by the rubrics. For example, the students are required to explain why the "third party" in the negative externalities don't have a say in the market. However, some students identified animals or the environment as the third party, making it unnecessary to explain why these parties lack a voice in the markets. Since these nuances are challenging to fully capture within rubrics, AI feedback that relies strictly on rubrics often fails to identify such edge cases.

Third, the TAs would also evaluate and provide additional feedback on the flow, conciseness and clarity of the essay, which are not required by the rubrics. For example, P2 advised a student to connect their solutions to the described scenario. P5 flagged a confusing sentence in the essay. P2, P3 and P4 all left additional comments on where the student could be more succinct. 

Fourth, TAs would emphasize what they want the student to learn when they provide feedback, and make sure that the feedback can be addressed by the content taught in the class. As P3 remarked, AI feedback didn't catch the student analysis being out of scope. P3 said \textit{"They were using concepts that weren't necessarily relevant to the scope of the assignment. But it's hard to kind of know that unless you know what's being taught in class already."} 
P5 also considered content covered in prior assignments when offering feedback, saying \textit{``The students did not have to (elaborate on) [a rubric], because at this point in the semester, I think they're able to identify that.''}
\vspace{-1pc}
\subsection{RQ2: What are the prospects of using AI-generated feedback as suggestions in the TA grading process? How should it be presented to enhance grading effectiveness?}

\subsubsection{Highlighting the relevant texts for each piece of AI feedback could speed up reading and assessments.}

Participants shared that the AI highlights in the Word document helped them more rapidly identify key ideas in students’ responses and thus sped up both reading and evaluation. As P4 said, \textit{``Figure[ing] out what the AI picked up... I think that would be faster than me having to read carefully through the essay twice.''}. Similarly, P3 noted that highlighting helps them understand the students' essays, and ensures important concepts are not overlooked. They said, \textit{``So I think with AI it's able to parse it out a little further to reduce the amount of time to help me understand. If I missed that they're getting the concept because they said it later on, it [AI highlight] would help me point that out and see that.''} 
Participants found the highlights especially useful when specific terms or evidence were required by the rubric, as they could confirm the presence of these terms at a glance. For example, P2 mentioned, \textit{``I sometimes feel like, I'm looking for specific points that they're hitting. And the AI highlights those specific points... So I definitely think it would help save time.''}
\vspace{-2pc}
\subsubsection{AI feedback could save TAs' time constructing feedback.}
Participants found it challenging to construct meaningful feedback for each student. For example, P4 said, \textit{``The main time is writing those comments cause you gotta write them individually for everyone.''} P2 further emphasized the difficulty in posing guiding questions in the feedback, \textit{``The hardest part is not giving them the answer, but also leading them in the right direction, because you can't be too explicit.''}
Participants shared that having AI feedback could facilitate their feedback writing. P4 said, \textit{``I think it's hard for me to write those comments because they can get a little bit wordy, but for the AI to do it, it's quick and it's easy and I think that's really good to use.''}. When participants found AI feedback that they could directly adopt, they applauded that it would save a lot of time. 
\vspace{-1pc}
\subsubsection{AI feedback could improve consistency in grading within one TA}

Several participants worried about the fairness of their own grading. As P1 explained, \textit{``I also worry about consistency. Like, if I take 1 point off for one student here. Did I take off 2 points for another student?''}. P5 also mentioned the same concern, wondering whether they might \textit{``grade certain students harsher than others (unintentionally).''} We did observe an inconsistency in TAs' grading when the student responses implied the correct idea but were imprecise. For example, for the rubric item, ``Mention that automation and labor are substitutes for consumption'', P4 was okay with one student using the word ``alternatives'' in one session, but decided to deduct points in a different session.
In such uncertain scenarios, participants often wanted a second opinion. P2 noted, \textit{``Sometimes we'll have questions like, is this acceptable? ... So it's nice to have something to consult.''} 
Several participants considered AI feedback to help them verify their evaluation, e.g., 
P2 said, \textit{``It's just like having another set of eyes on the paper.''} 

Moreover, several participants shared that the AI feedback made them think a little harder about the rubrics. For instance, both P4 and P5 realized they had overlooked certain rubric criteria in their evaluations, which the AI had identified.
P5 commented after reading the AI feedback for several students: \textit{``I’m reading through these (AI) comments. And I’m like, Wow, I feel like I’m not catching a lot of things.''}. Consequently, they found AI feedback \textit{``helpful to have standardization within your own section''}
\vspace{-1pc}
\subsubsection{AI feedback could help standardize grading among TAs}
Although instructors developed the rubrics and held regular staff meetings throughout the semester to help TAs learn and apply them consistently, TAs still observed inconsistency in their application and expressed concerns about this issue. P5 said \textit{``I think something that we are always concerned about is that one [TA] is grading too leniently versus other [TA] that's grading harshly.''} Both P1 and P2 noticed that they were more lenient on certain rubrics than other TAs. One main source of inconsistencies lies in the varying interpretations of the rubrics. Some rubrics defined aspects that students should address, and there remains flexibility in the detailedness and depth of analysis. For example, one rubric item requires ``a thoughtful and well-reasoned solution'', which leaves significant discretion to TAs. P5 expected a thorough explanation of why a solution would be effective, whereas P2 found that a correctly named potential solution was sufficient. Moreover, even when the idea is specified in the rubric, the judgment on students' alternative ways to express the idea also leads to inconsistency.
Participants envision that AI feedback could help build consistency in grading among TAs.

\vspace{-1pc}
\subsubsection{Concern about over-relying on AI assistance}
Participants are aware of potential AI mistakes. They also shared concerns about missing key points and making mistakes if they fully rely on AI. For example, P3 found reading AI feedback in the middle of reading the student's essay distracting, and P4 shared concerns about overlooking sentences that were not highlighted. Almost all participants mentioned that to ensure all the important points were covered and all the errors were caught, they would read the essay and make their own judgment first, before they read and evaluate the AI judgment and feedback. As P5 said, \textit{``You don't want to like over-rely on it (AI), in case something is inconsistent.''}

\vspace{-1.25pc}
\section{Discussion, Limitation and Design Implications}
\vspace{-0.75pc}
Our study highlights both the strengths and limitations of AI-generated feedback. While AI feedback offers benefits such as stronger alignment with rubrics and detailed, personalized explanations, it may contain errors and ineffective messages. Using AI feedback as a supplementary tool to enhance TAs’ grading processes shows significant promise. In this section, we discuss three key considerations when designing human-AI collaborative systems to support the process: 1) Establish clearly written rubrics; 2) Use highlights to increase transparency on how feedback is generated; 3) Provide intermediate outputs on each AI subtask, so that users can decide which outcomes to consume.

\textbf{Establish clearly written rubrics.} Our study highlights the importance of rubrics for AI to generate high quality feedback.
We found that further explanation and clarification are needed beyond the original rubrics TAs use for the LLM to make accurate evaluations, in line with \cite{wu2024unveiling}. We provide tips in Table \ref{tab:rubric} on elaborating rubrics to make them more understandable by LLMs.
\begin{table}[]
\vspace{-1pc}
\caption{Suggestions for elaborating rubrics in order for LLMs to generate accurate feedback that is aligned with expectations for knowledge-intensive essays.}
\vspace{-0.5pc}
\begin{tabular}{|l|l|l|}
\hline
   & Good rubric example & Bad rubric example                                                                                                                      \\ \hline
\begin{tabular}[c]{@{}l@{}}Explain the \\ domain-specific \\ knowledge\end{tabular}                                                         & \begin{tabular}[c]{@{}l@{}}The student demonstrated an understanding \\ of the Law of Demand, that is, as the price \\ of the good increases, the quantity \\ demanded by the good or service decreases.\end{tabular} & \begin{tabular}[c]{@{}l@{}}The student correctly \\ used the terms quantity \\ supplied/demanded vs. \\ supply/demand.\end{tabular}     \\ \hline
\begin{tabular}[c]{@{}l@{}}Include acceptable \\ alternatives\end{tabular}                                                                   & \begin{tabular}[c]{@{}l@{}}The student demonstrated that farmers \\ demand water, or analyze the influence on \\ farmers as consumers of water.\end{tabular}                                                          & \begin{tabular}[c]{@{}l@{}}The student stated that \\ with tax on\\ automation,the demand \\for labor increases.\end{tabular} \\ \hline
\begin{tabular}[c]{@{}l@{}}Specify the \\ expected depth of \\ the explanation\end{tabular}                                                  & \begin{tabular}[c]{@{}l@{}}The student explained why deadweight loss \\ exists and mention it is quite large given \\ that the Government purchased the excess.\end{tabular}                                          & \begin{tabular}[c]{@{}l@{}}Explain the concept of \\ artificially scarce goods \\ conceptually.\end{tabular}                            \\ \hline
\begin{tabular}[c]{@{}l@{}}Negative behaviors \\should be explicity\\called out\end{tabular} & \begin{tabular}[c]{@{}l@{}}The student did not use long direct quotes \\ (more than 1 sentence in one quote) from \\ the article.\end{tabular}                                                                         & \begin{tabular}[c]{@{}l@{}}Direct in-text references \\ are present.\end{tabular}                                                       \\ \hline
\end{tabular}
\label{tab:rubric}
\vspace{-1.5pc}
\end{table}

\textbf{Use highlights to increase transparency on how feedback is generated.}
Many participants shared that even when AI makes mistakes in determining whether a rubric item is satisfied, the highlighted sentences in the essay made it easy for them to understand and correct AI's judgment. 
For example, P4 said, \textit{``even if it's sometimes incorrect, that's what you can check ... I think that's the easiest part for us to get.''} After grading 2 essays, P4 found a pattern in what rubric AI might make mistakes. They would be more careful and mainly relied on their own judgments.
On the other hand, they appreciate AI for providing detailed and personalized feedback and giving more praise, as well as guiding questions, which are time-consuming for the TAs to write themselves. Our findings indicate that highlighting the sentences used for the AI outputs is an effective way to visualized the AI's rationales without requiring users to read additional explanatory text.

\textbf{Provide intermediate outputs on each AI subtask, so that users can decide which outcomes to consume.}
Our findings showed that the three-step feedback engine was effective in generating high quality feedback for knowledge-intensive essays. For each rubric, the three steps include 1) identify relevant sentences in the essay for this rubric; 2) make a judgment on whether this rubric is satisfied or not; 3) generate a feedback message. 
This decomposition of the feedback generation task gives users the flexibility to decide whether they should take the AI results. For example, AI might make mistakes on the second step, i.e., evaluating whether the rubric is satisfied. With the highlighted sentence, the user can locate the mistake more easily and flip the judgment. AI might also generate suboptimal feedback, e.g., revealing the correct answer. With the highlighted sentences and the suggested historic and AI feedback, users can decide to either use the historic feedback, write their own feedback, or combine both feedback to craft a better message. 
Our findings suggest that participants were aware of AI’s potential hallucinations and inaccuracies, and the visibility of intermediate outputs helped them better evaluate the correctness of AI suggestions. 

In this work, we concentrated on knowledge-intensive essays. Such essay assignments are widely adopted in STEM courses to support student learning in various domains, including Materials Science
and Engineering, Organic Chemistry, Introductory
Statistics, etc. \cite{finkenstaedt2023portrait} Our work looks into the TA's perspective and highlights the promising prospects of AI-generated feedback. TAs found AI feedback more detailed and aligned with the rubrics, while they usually adopt a more wholistic approach. Future work could explore whether such point-to-point feedback is preferable. Moreover, as our findings show positive prospects of AI feedback, future work could further investigate how  feedback created with AI feedback as suggestions would impact students' learning outcomes.  


\vspace{-1pc}
\section{Conclusion}
\vspace{-0.5pc}
This study aims to examine the prospect of using AI-generated feedback as suggestions to expedite and enhance human instructors' feedback provision. We performed the study in a college-level introductory economics class, which often assigns knowledge-intensive essays, where students need to demonstrate a precise understanding of economic concepts in their writing. Despite providing detailed and clearly-written rubrics, and using a carefully designed feedback engine, there are inevitable hallucinations, mistakes, and ineffectiveness in AI feedback. This study argues for a future where AI feedback can be provided as suggestions during human instructors' grading and feedback provision process. Participants responded positively to the use of AI suggestions and perceived them to accelerate grading and enhance feedback quality. Moreover, we highlight the importance of providing detailed and comprehensive rubrics when using AI to provide feedback on knowledge-intensive essays.

\begin{credits}
\subsubsection{\ackname} This work was funded by NSF Grants IIS-2302564. The findings and conclusions expressed in this material are those of the author(s) and do not necessarily reflect the views of the National Science Foundation.

\end{credits}
%
%
%
\bibliographystyle{splncs04}
\bibliography{reference}
%




\end{document}